# Spin blockade and coherent dynamics of high-spin states in a three-electron double quantum dot

Bao-Bao Chen,[1,2,*] Bao-Chuan Wang,[1,2,*] Gang Cao,[1,2,†] Hai-Ou Li,[1,2] Ming Xiao,[1,2] Guang-Can Guo,[1,2] Hong-Wen Jiang,[3] Xuedong Hu,[4] and Guo-Ping Guo[1,2,†]

[1] Key Laboratory of Quantum Information, University of Science and Technology of China,
Chinese Academy of Sciences, Hefei 230026, China
[2] Synergetic Innovation Center of Quantum Information & Quantum Physics,
University of Science and Technology of China, Hefei, Anhui 230026, China
[3] Department of Physics and Astronomy, University of California at Los Angeles, California 90095, USA
[4] Department of Physics, University at Buffalo, SUNY, Buffalo, New York 14260, USA

*These authors contributed equally to this work
†Corresponding Authors: gpguo@ustc.edu.cn; gcao@ustc.edu.cn

## Abstract

Asymmetry in a three-electron double quantum dot (DQD) allows spin blockade, when spin-3/2 (quadruplet) states and spin-1/2 (doublet) states have different charge configurations. We have observed this DQD spin blockade near the (1,2)-(2,1) charge transition using a pulsed-gate technique and a charge sensor. We then use this spin blockade to detect Landau-Zener-Stückelberg (LZS) interference and coherent oscillations between the spin quadruplet and doublet states. Such studies add to our understandings of coherence and control properties of three-spin states in a double dot, which in turn would benefit the explorations into various qubit encoding schemes in semiconductor nanostructures.

Spins in semiconductor quantum dots are a promising candidate for qubit [1]. Experimental studies of spin qubits have overcome a multitude of obstacles in the past decade in demonstrating spin preparation, manipulation, and measurement. A key enabling technology for these achievements is single spin detection via spin blockade [2], which allows spin states to be distinguished by a charge sensor quickly [3].

While single-electron spin is an obvious candidate for a qubit [4-12], encoding in multi-electron states [3,13-26] could remedy the main shortcoming of a single-spin qubit, i.e. the very slow single-qubit gate. Indeed, encoded three-spin qubits, whether in a triple dot [16-21] or a double dot [22-26], can be controlled completely electrically. Potential weakness of such encoded qubits lies in the more complex state structure, possible leakage [27], and susceptibility to charge noise [20,28,29]. It is therefore important to acquire a comprehensive understanding of the spectrum and dynamics of a multi-electron system in order to establish its feasibility for qubit encoding.

In this letter we investigate three-electron spin dynamics in a GaAs double quantum dot (DQD). We first identify how spin blockade between electron spin doublet and quadruplet states can be realized in a three-electron double dot, and experimentally verify its existence. We then study LZS interference and coherent oscillations between doublet and quadruplet states near the (2,1)-(1,2) transition of the DQD. Our results provide important insights into the coherence and controllability of multi-electron states in a DQD.

The double dot device used in this letter is fabricated on a $GaAs/Al_{0.3}Ga_{0.7}As$ heterostructure, as shown in Fig. 1(a). Charge configuration of the double dot is monitored by measuring transconductance $dI_{QPC}/dV_R$ of a nearby quantum point contact (QPC). High-frequency pulses are applied on gates $L$ and $R$ to change the inter-dot detuning rapidly. The device is placed in a dilution refrigerator with a base temperature of 30 mK. An external magnetic field is applied in-plane, and is perpendicular to the axis connecting the two dots. The double dot is operated near the (1,2)-(2,1) charge transition, as shown in Fig. 1(b), where (N,M) denotes electron numbers on the left and right dot.

The low-energy spectrum of the three-electron DQD consists of two spin doublets and one quadruplet [18,30]. In the (1,2) [(2,1)] charge configuration, spin doublet state $D_S(1,2)$ [$D_S(2,1)$] with $S=1/2$ is the ground state, with the two electrons in the right (left) dot forming a singlet state. When the two electrons in the right (left) dot form a triplet state, the three-electron states of the DQD consist of a spin quadruplet $Q(1,2)$ [$Q(2,1)$] with $S=3/2$ and another spin doublet $D_T(1,2)$ [$D_T(2,1)$] with $S=1/2$. They are $E_R$ ($E_L$) higher in energy as compared to $D_S(1,2)$ [$D_S(2,1)$] [Fig. 1(c)]. Here $E_R$ ($E_L$) is the singlet-triplet splitting in the right (left) dot, and is closely and inversely related to the dot size. In our system, the left dot is smaller than the right dot, so that $E_L > E_R$. When the inter-dot detuning is varied, as shown in Fig. 1(d), a series of anti-crossings develop, mostly due to spin-independent tunnel coupling. Examples include those between $D_S(2,1)$ and $D_S(1,2)$ at $\varepsilon=0$, $D_S(2,1)$ and $D_T(1,2)$ at $\varepsilon=-E_R$, and $Q(1,2)$ and $Q(2,1)$ at $\varepsilon=E_L-E_R$. The splittings of the first two anti-crossings are $2\Delta_1$ and $2\Delta_2$, respectively. Usually $\Delta_2 \gg \Delta_1$ since $D_T(1,2)$ involves more extended excited orbital state.

A magnetic field $B$ splits sublevels of a spin multiplet by Zeeman energy $E_Z = g\mu_B B$, as shown in Fig. 1(e). Here $g$ is the electron gyromagnetic ratio and $\mu_B$ is the Bohr magneton. Two anti-crossings are of particular interest to us. One is between $D_S(2,1)_{+1/2}$ and $D_T(1,2)_{+1/2}$ due to spin-independent tunneling. We use $\tilde{D}_{+1/2}$ to denote the lower branch in the negative detuning region, as the two states mix. As $\varepsilon$ decreases, $\tilde{D}_{+1/2}$ approaches degeneracy with $Q(1,2)_{+1/2}$ in the absence of nuclear field. From here on we use $J(\varepsilon)$ to denote the energy splitting between $Q(1,2)_{+1/2}$ and $\tilde{D}_{+1/2}$. Another anti-crossing is between $Q(1,2)_{+3/2}$ and $\tilde{D}_{+1/2}$ at detuning $\varepsilon_I$ [see inset of Fig. 1(e)]. This small anti-crossing is due to spin-flip tunneling enabled by the nuclear Overhauser field in GaAs [31,32]. We need this anti-crossing to populate the $Q(1,2)_{+3/2}$ state from the ground $D_S(2,1)_{+1/2}$ state.

Spin blockade refers to parameter regimes in which states with different spin symmetries also have different charge configurations [33]. This spin-charge correlation leads to current rectification through a double dot [34,35], while also allows detection of spin states via charge sensing. A close inspection of Fig. 1(d) shows that in and only in the detuning window $0 < \varepsilon < E_L - E_R$, we indeed have such a spin-charge correlation. Specifically, here the ground doublet state is $D_S(2,1)$, while the ground quadruplet state is $Q(1,2)$. Together with a strong suppression of spin-flip tunneling, a spin blockade window is established.

To experimentally demonstrate the blockade of $Q$ states, we initialize into the ground $D_S(2,1)_{+1/2}$ state, then apply a simple square detuning pulse (toward the negative detuning direction) as shown in Fig. 2(a) [36]. The rise and fall of the pulse is sufficiently fast so that the system passes through the $D_S - D_S$ and $\tilde{D}_{+1/2} - Q_{+3/2}$ anti-crossings diabatically. If the pulse height is such that the double dot is detuned to the $\tilde{D}_{+1/2} - Q_{+3/2}$ anti-crossing at the pulse tip, though, state mixing occurs and $Q(1,2)_{+3/2}$ is populated. When the system is then pulsed back into the spin blockade regime, $Q(1,2)_{+3/2}$ remains trapped in the (1,2) charge configuration because its relaxation involves spin flip, and can be detected by the charge sensor.

Figure 2(b) shows $dI_{QPC}/dV_R$ as a function of the base detuning $\varepsilon_b$ and amplitude $V_P$ of the pulse. The V-shaped region enclosed by the white dashed lines in Fig. 2(b) is where the system can pass through the $D_S - D_S$ anti-crossing driven by the pulse sequence we employ. During the rising edge of the pulse, the system passes through $D_S - D_S$ anti-crossing diabatically since $\Delta_1$ is tuned sufficiently small. Thus the system populates $\tilde{D}_{+1/2}$ with high probability. If the pulse tip reaches the $\tilde{D}_{+1/2} - Q_{+3/2}$ anti-crossing, $Q(1,2)_{+3/2}$ is populated and is trapped in (1,2) charge state after detuning is pulsed back into the spin blockade window. This signal can be picked up by the charge sensor and shows up as a resonance line [the green solid line in Fig. 2(b)]. By varying the applied field $B$, we can track the position of the $\tilde{D}_{+1/2} - Q_{+3/2}$ anti-crossing and obtain the characteristic spin funnel [Fig. 2(d)], which provides a direct measurement of $J(\varepsilon)$ [3]. A transition line [the red solid line in Fig. 2(b)] truncates the resonance line, marking where spin blockade is lifted. This means the base detuning has reached the anti-crossing between $Q(1,2)$ and $Q(2,1)$, which yields $E_L - E_R = 1.32$ meV. Lastly, if the voltage pulse is applied along the positive detuning direction, spin blockade is not observed in the corresponding V-shaped region in Fig. 2(c), illustrating the directional feature of spin blockade.

Having established the *Q*-state spin blockade, we can now study coherent dynamics between *Q* and *D* states. In the remainder of this letter, the measurement point $\varepsilon_M$ is always set in the blockade window [Fig. 1(e)].

The first experiment we perform is to generate LZS interference [31,32] by sweeping the detuning past the $\tilde{D}_{+1/2} - Q_{+3/2}$ anti-crossing. For this experiment the system is initialized to the ground state $D_S(2,1)_{+1/2}$ in the positive detuning region. We apply a specially designed detuning pulse as shown in Fig. 3(a), with multiple values of rising speed. Specifically, we set pulse parameters $t_{Dr}$ and $t_{Df}$ as small as we can, with the rise time (~1 ns) limited by the bandwidth of the transmission line. The fast rise allows the initial $D_S(2,1)_{+1/2}$ state to pass through the $D_S - D_S$ anti-crossing diabatically and becomes a $\tilde{D}_{+1/2}$ state. We set $t_U = 5$ ns to make a slower passage through the $\tilde{D}_{+1/2} - Q_{+3/2}$ anti-crossing, where $\tilde{D}_{+1/2}$ evolves into a superposition of $\tilde{D}_{+1/2}$ and $Q(1,2)_{+3/2}$ states. The system is then held at a detuning $\varepsilon < \varepsilon_I$ for a time period $\tau_S$, resulting in a phase accumulation $\phi$ between the two components of the superposition. A reverse ramp then takes the system back across the $\tilde{D}_{+1/2} - Q_{+3/2}$ anti-crossing where interference occurs. At the end of the pulse, the system is taken back to the measurement point for spin readout.

Figures 3(b) and 3(c) present the QPC conductance $dI_{QPC}/dV_R$ as a function of $\varepsilon$ and $\tau_S$, measured at $B = 110$ mT and $B = 60$ mT respectively. Both display typical LZS interference patterns [31,32]. The oscillation frequency is set by energy difference between $\tilde{D}_{+1/2}$ and $Q(1,2)_{+3/2}$ at $\varepsilon$, where phase $\phi$ is acquired. This frequency $[E_Z - J(\varepsilon)]/h$ increases with the applied field $B$, as we expect.

In Fig. 3(d), we use a Gaussian damped cosine function to fit a trace [red line in Fig. 3(d)] taken from Fig. 3(b) at the detuning marked by the red dashed line, and extract a dephasing time of $T_2^* = 7.0$ ns. The dephasing is most likely a result of fluctuations in the nuclear field [31,32], which determines the energy splitting of the $\tilde{D}_{+1/2} - Q_{+3/2}$ anti-crossing and therefore how the population is split between the two states during the detuning sweep.

One key to the LZS experiment at the $\tilde{D}_{+1/2} - Q_{+3/2}$ anti-crossing is that the initial $D_S(2,1)_{+1/2}$ state passes through the $D_S - D_S$ anti-crossing diabatically and populates the $\tilde{D}_{+1/2}$ state. Here we perform two control experiments to clarify the effects of the $D_S - D_S$ anti-crossing. We first set $t_{Dr} = 5$ ns and $t_{Df} = 0$ to make the first passage across $D_S - D_S$ anti-crossing adiabatic. As expected, oscillation disappears [bottom blue line in Fig. 3(d)] because the system cannot be initialized into $\tilde{D}_{+1/2}$ efficiently. Alternatively, we set $t_{Dr} = 0$ and $t_{Df} = 5$ ns to make the last passage across $D_S - D_S$ anti-crossing adiabatic. Now the amplitude of the oscillation is almost unaffected [middle green line in Fig. 3(d)], because the last passage only drives transitions between the doublet states, and $D_S(1,2)_{+1/2}$ relaxes to $D_S(2,1)_{+1/2}$ rapidly. Thus the final readout remains unchanged. These two control experiments show that even though the DQD has to pass through the $D_S - D_S$ anti-crossing, our pulse design does allow the QPC readout to give us reliable information on what happens at the $\tilde{D}_{+1/2} - Q_{+3/2}$ anti-crossing.

One interesting sector of the energy spectrum shown in Fig. 1(e) is the subspace spanned by $\tilde{D}_{+1/2}$ and $Q(1,2)_{+1/2}$, which is remarkably similar to that spanned by $S$ and $T_0$ in a two-electron DQD [3]. For example, we can perform an exchange operation in this subspace by employing the pulse sequence shown in Fig. 4(a), similar to what was used in Ref. [3]. Here the system is initialized into $D_S(2,1)_{+1/2}$ at a positive detuning. After diabatic passages across the $D_S - D_S$ anti-crossing and the $\tilde{D}_{+1/2} - Q_{+3/2}$ anti-crossing consecutively, $\tilde{D}_{+1/2}$ is prepared. The detuning is then ramped adiabatically to a point where $J$ is small, so that $\tilde{D}_{+1/2}$ approaches the ground state of the nuclear field, $|\downarrow\rangle|T_+\rangle$ or $|\uparrow\rangle|T_0\rangle$ [30], depending on the instantaneous nuclear field across the double dot. On the Bloch sphere [Fig. 4(b)] where the $z$ axis is defined by $\tilde{D}_{+1/2}$ and $Q(1,2)_{+1/2}$, our eigenstates define a tilted axis relative to $z$ [14]. After initialization, an exchange pulse (a sudden change of detuning toward $\varepsilon_I$) switches on $J(\varepsilon)$ for a time period of $\tau_S$, which rotates the state vector around the $z$ axis on the Bloch sphere with a frequency $J(\varepsilon)/h$. The final spin state is read out by a reverse adiabatic ramp which maps $|\downarrow\rangle|T_+\rangle$ [$|\uparrow\rangle|T_0\rangle$] to $\tilde{D}_{+1/2}$ [$Q(1,2)_{+1/2}$].

In Fig. 4(c) we plot $dI_{QPC}/dV_R$ as a function of $\varepsilon$ and $\tau_S$. The conductance indeed shows an oscillation whose frequency increases with $\varepsilon$. Fast Fourier transform (FFT) over $\tau_S$ of the data in Fig. 4(c) gives the frequency of the oscillations, and is shown in Fig. 4(d). It exhibits a funnel shape similar to that in Fig. 2(d). Peak centers at each detuning [green circles in Fig. 4(d)] give $J(\varepsilon)$. A simple model for $J(\varepsilon)$ [30], which includes a phenomenological suppression of tunnel coupling [17,19], fits well with the data [red curve in Fig. 4(d)], yielding parameters $E_R = 0.27$ meV and $\Delta_2 = 21$ μeV. From previously obtained $E_L - E_R$ we get $E_L = 1.59$ meV, significantly different from $E_R$, confirming our assertion on the asymmetry of the DQD. Comparing $J(\varepsilon)$ obtained here with spin funnel in Fig. 2(d), we find good agreement if $g = -0.35$ [red curves in Fig. 2(d)].

In conclusion, we identify and observe spin blockade of spin quadruplet states relative to spin doublet states in a three-electron GaAs double quantum dot. With the help of spin blockade, and using pulsed-gate technique with specially designed pulse sequences based on the electron spectrum, we are able to study the coherent dynamics between spin quadruplet and doublet states. Specifically, we have studied LZS interference near the anti-crossing between the $S_z = +1/2$ doublet and the $S_z = +3/2$ quadruplet states, and examined the three-electron dynamics in the subspace spanned by the $S_z = +1/2$ quadruplet state and the $S_z = +1/2$ doublet state. Our results reveal that the new type of spin blockade could be a powerful tool for future investigations in multi-electron quantum dots, and pulsed-gate technique is an insightful substitute for transport experiments in studying multi-electron spectrum and dynamics in a double dot.

**Acknowledgements:** This work was supported by the National Key R & D Program (Grant No.2016YFA0301700), the NNSF (Grant Nos. 11304301, 11575172, 61306150, and 91421303), the SPRP of CAS (Grant No. XDB01030000), and the Fundamental Research Fund for the Central Universities. XH and HWJ acknowledge financial support by US ARO through grant W911NF1210609 and W911NF1410346, respectively.

# FIGURE CAPTIONS

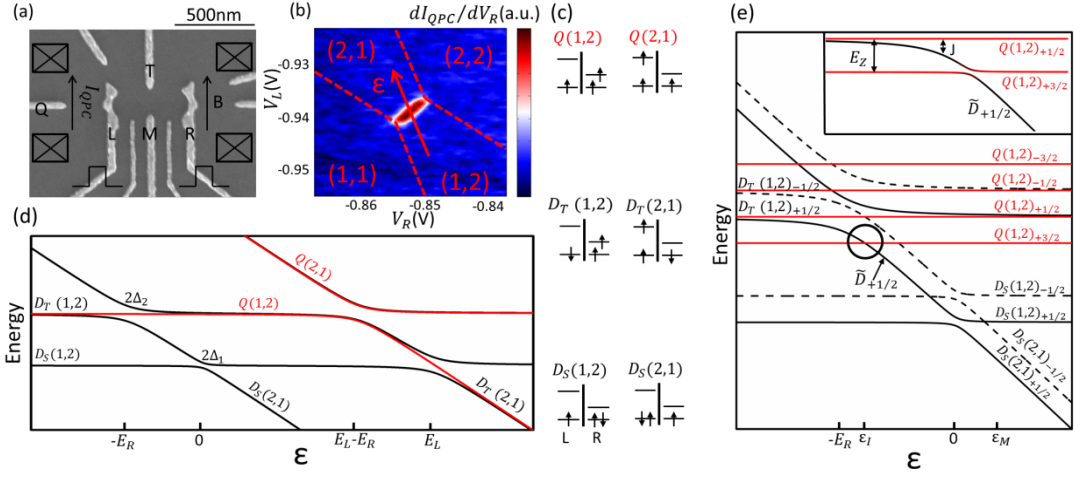

**FIG. 1.** (a) Scanning electron microscope image of a double dot device. (b) Stability diagram near (1,2)-(2,1) charge transition with charge numbers in different regions labeled. Red dashed lines represent missing charge addition lines due to small tunnel rates between dots and reservoirs. Red arrow defines the detuning axis. Two synchronized pulses on gate $L$ and $R$ are calibrated to change detuning along this axis. (c) Schematics of electron fillings for different spin and charge states. Only the two lowest orbitals in both dots and configurations with lowest possible energy are considered. (d) Energy levels as a function of detuning for $B=0$. (e) Energy diagram at the left of $\varepsilon \sim E_L - E_R$ in (d) for $B>0$. Inset: zoom in plot of the portion near the black circle, highlighting states used for coherent manipulation and detailing the anti-crossing between $\tilde{D}_{+1/2}$ and $Q(1,2)_{+3/2}$.

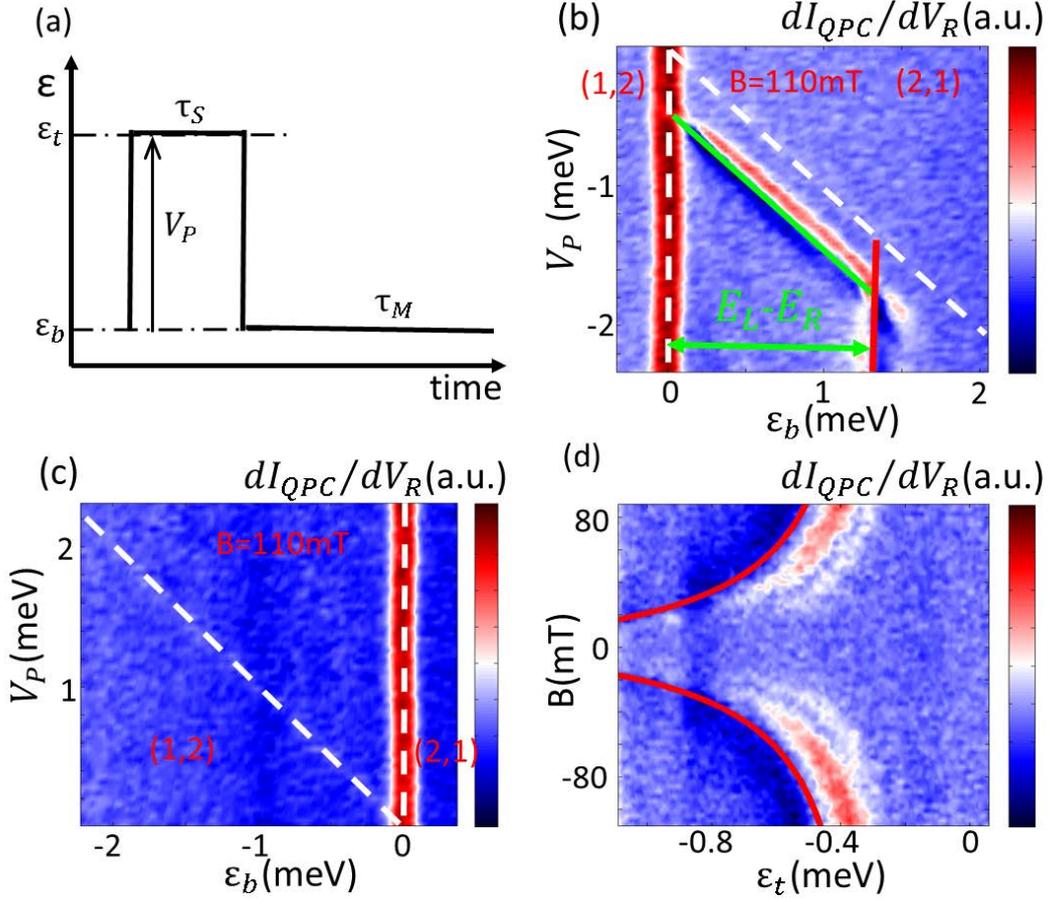

**FIG. 2.** (a) Detuning pulse used to observe spin blockade and spin funnel with $\tau_S = 200$ ns and $\tau_M = 5\,\mu\text{s}$. (b) $dI_{QPC}/dV_R$ measured at $B = 110$ mT with a negative detuning pulse, as a function of base detuning $\varepsilon_b$ and pulse amplitude $V_P$. (c) Same as in (b) but with the pulse applied along positive detuning direction. (d) $dI_{QPC}/dV_R$ measured with a negative detuning pulse, as a function of detuning of the pulse tip, $\varepsilon_t$ and magnetic field $B$, exhibiting a funnel-shaped feature. The base detuning $\varepsilon_b$ is set in the blockade window. Red curves are calculated spin funnel using the model for $J(\varepsilon)$ and a best-fit $g = -0.35$.

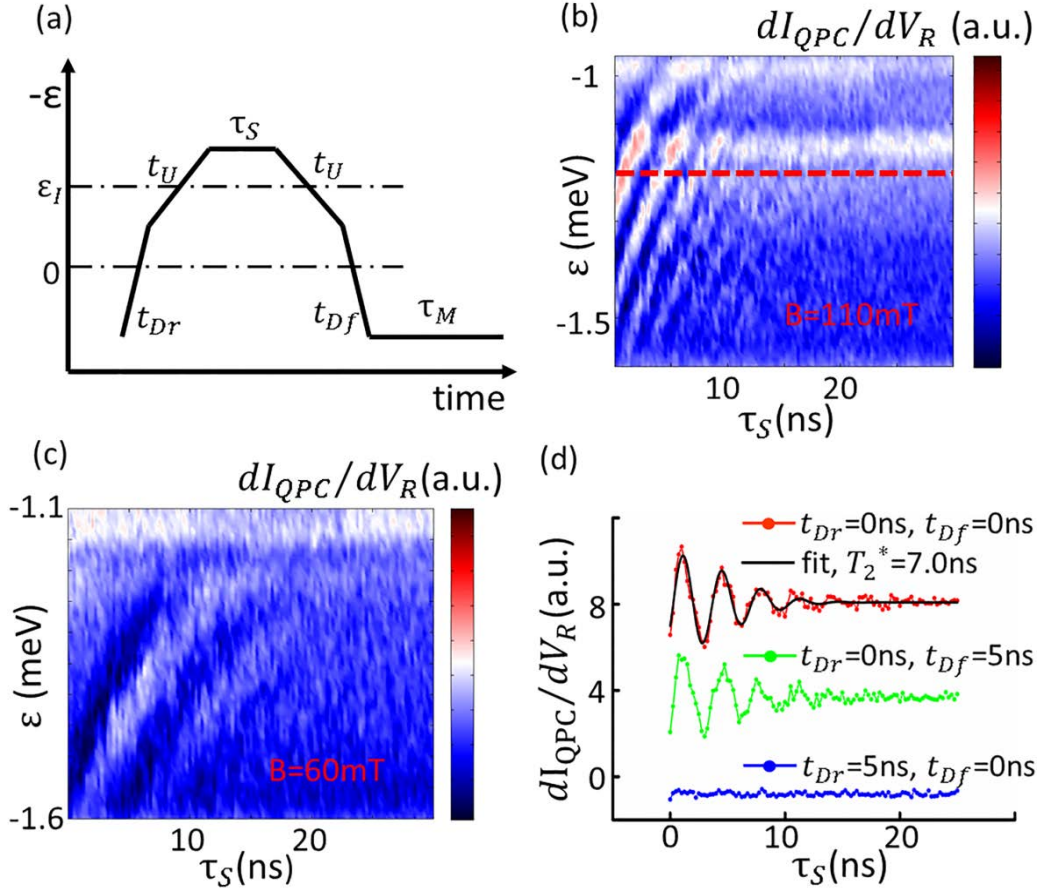

**FIG. 3.** (a) Detuning pulse used to measure LZ oscillations. (b), (c) LZ oscillations as a function of $\varepsilon$ and $\tau_S$ measured at $B = 110$ mT (b) and $B = 60$ mT (c). (d) Traces measured at the detuning marked by red dashed line in (b), with different pulse parameters related to $D_S - D_S$ anti-crossing. $t_U$ is set to 5 ns for all the cases. Fitting top red line to a Gaussian damped cosine function yields a $T_2^* = 7.0$ ns.

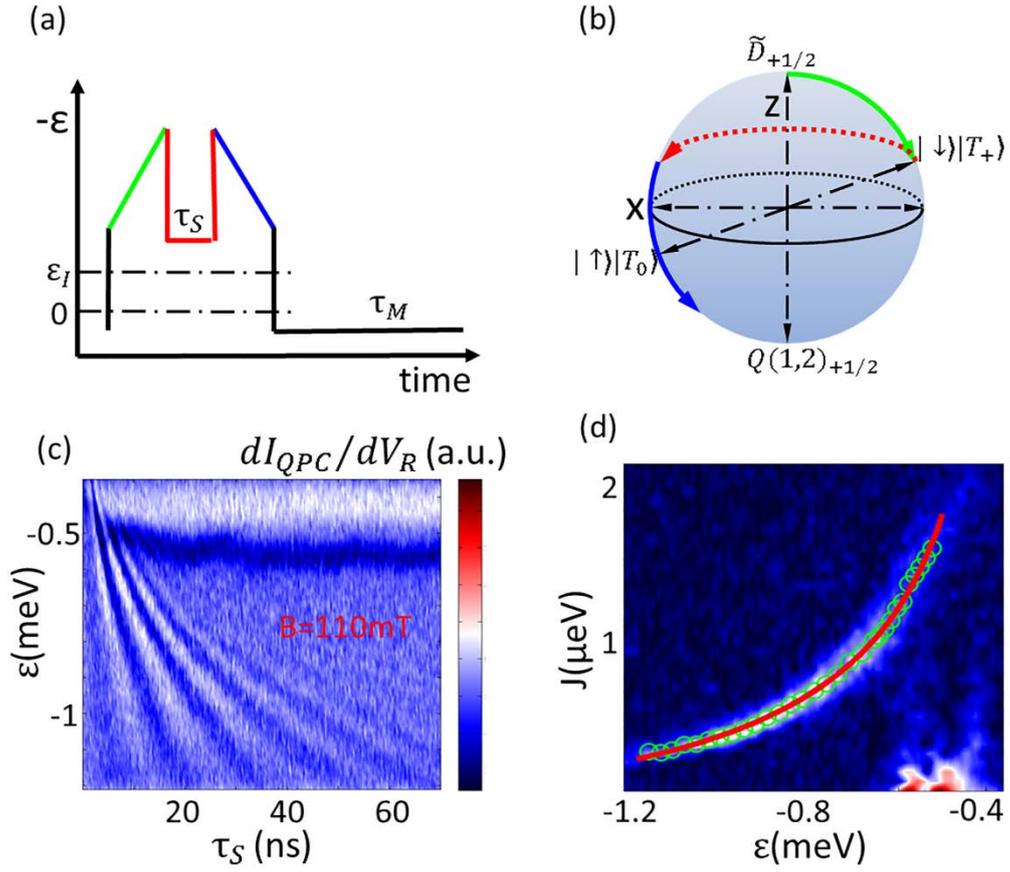

**FIG. 4.** (a) Detuning pulse used to measure exchange oscillations. (b) Bloch sphere representation of state vector evolutions under corresponding pulse stages. (c) Exchange oscillations measured at $B = 110\,\text{mT}$, as a function of $\varepsilon$ and $\tau_S$. (d) FFT of data in (c) over $\tau_S$. The vertical axis has been converted from frequency to energy by multiplying Planck's constant. Green circles are peak centers found by Gaussian fit. Red curve is a fit to a model for $J(\varepsilon)$.

# Supplementary Information for Spin Blockade and Coherent Dynamics of High-Spin States in A Three-Electron Double Quantum Dot

Bao-Bao Chen,[1, 2, *] Bao-Chuan Wang,[1, 2, *] Gang Cao,[1, 2, †] Hai-Ou Li ,[1, 2] Ming Xiao,[1, 2] Guang-Can Guo,[1, 2] Hong-Wen Jiang,[3] Xuedong Hu,[4] and Guo-Ping Guo[1, 2, †]

[1] Key Laboratory of Quantum Information, University of Science and Technology of China,
Chinese Academy of Sciences, Hefei 230026, China
[2] Synergetic Innovation Center of Quantum Information & Quantum Physics,
University of Science and Technology of China, Hefei, Anhui 230026, China
[3] Department of Physics and Astronomy, University of California at Los Angeles, California 90095, USA
[4] Department of Physics, University at Buffalo, SUNY, Buffalo, New York 14260, USA

*These authors contributed equally to this work
†Corresponding Authors: gpguo@ustc.edu.cn; gcao@ustc.edu.cn

## I. spin states

The low energy sector of a three-electron Hilbert space contains eight independent spin states [1], including one quadruplet $|Q_{S_z}\rangle$ with total spin $S=3/2$ and z-component $S_z=\pm 3/2, \pm 1/2$, and two doublets, $|D_{SS_z}\rangle$ and $|D_{TS_z}\rangle$ with $S=1/2$ and $S_z=\pm 1/2$. In a double dot in the (1,2) [(2,1)] charge configuration, the two doublets are distinguished by the symmetry of their constituencies in the two-electron right (left) dot. The explicit expressions for these eight states in the (1,2) configuration are:

$$|Q_{+3/2}\rangle = |\uparrow\rangle |T_+\rangle, \tag{S1}$$

$$|Q_{+1/2}\rangle = \sqrt{\frac{2}{3}}|\uparrow\rangle |T_0\rangle + \sqrt{\frac{1}{3}}|\downarrow\rangle |T_+\rangle, \tag{S2}$$

$$|Q_{-1/2}\rangle = \sqrt{\frac{2}{3}}|\downarrow\rangle |T_0\rangle + \sqrt{\frac{1}{3}}|\uparrow\rangle |T_-\rangle, \tag{S3}$$

$$|Q_{-3/2}\rangle = |\downarrow\rangle|T_-\rangle, \qquad (S4)$$

$$|D_{T+1/2}\rangle = \sqrt{\frac{1}{3}}|\uparrow\rangle|T_0\rangle - \sqrt{\frac{2}{3}}|\downarrow\rangle|T_+\rangle, \qquad (S5)$$

$$|D_{T-1/2}\rangle = \sqrt{\frac{1}{3}}|\downarrow\rangle|T_0\rangle - \sqrt{\frac{2}{3}}|\uparrow\rangle|T_-\rangle, \qquad (S6)$$

$$|D_{S+1/2}\rangle = |\uparrow\rangle|S\rangle, \qquad (S7)$$

$$|D_{S-1/2}\rangle = |\downarrow\rangle|S\rangle. \qquad (S8)$$

Here $|S\rangle$, $|T_+\rangle$, $|T_0\rangle$ and $|T_-\rangle$ are the spin states of the two-electron right dot, while $|\uparrow\rangle$ and $|\downarrow\rangle$ are the spin states of the single electron in the left dot.

## II. Model for $J(\varepsilon)$

The energy splitting $J(\varepsilon)$ between $Q(1,2)_{+1/2}$ and $\tilde{D}_{+1/2}$ is a result of level repulsion at the $D_S - D_T$ anti-crossing. This is similar to the two-electron case except for a difference in the position of anti-crossing [2]. The expression of $J(\varepsilon)$ can thus be obtained by shifting the corresponding one for the two-electron case [2] by $-E_R$ along the detuning axis:

$$J(\varepsilon) = \frac{\varepsilon + E_R}{2} + \sqrt{\frac{(\varepsilon + E_R)^2}{4} + \left\{\Delta_2 \exp\left[-(w(\varepsilon + E_R))^2\right]\right\}^2}. \qquad (S9)$$

Here we have included a phenomenological suppression of tunnel coupling by detuning [3,4], with a free parameter $w$ reflecting the strength.

## III. Eigenstates of the nuclear fields

Let us consider the sub-space spanned by $Q(1,2)_{+1/2}$ and $\tilde{D}_{+1/2}$ states deeply in the negative detuning region, where $J(\varepsilon) \approx 0$. Here the three-electron eigenstates are no longer $Q(1,2)_{+1/2}$ and $\tilde{D}_{+1/2}$, but are determined by the nuclear spin

Overhauser fields in the double dot. On the basis of $Q(1,2)_{+1/2}$ and $\tilde{D}_{+1/2}$, the hyperfine Hamiltonian is:

$$H_{hf} = g\mu_B \begin{pmatrix} \dfrac{B_{Lg}^z + B_{Rg}^z + B_{Re}^z}{6} & \dfrac{2B_{Lg}^z - B_{Rg}^z - B_{Re}^z}{3\sqrt{2}} \\ \dfrac{2B_{Lg}^z - B_{Rg}^z - B_{Re}^z}{3\sqrt{2}} & \dfrac{-B_{Lg}^z + 2B_{Rg}^z + 2B_{Re}^z}{6} \end{pmatrix}. \quad (S10)$$

Here $B_{Lg}^z$, $B_{Rg}^z$ and $B_{Re}^z$ are the z-components of the Overhauser fields experienced by an electron occupying the ground orbital in the left dot, the ground orbital in the right dot, and the first excited orbital in the right dot, respectively. The eigenstates of this hyperfine Hamiltonian are $|\downarrow\rangle|T_+\rangle$ and $|\uparrow\rangle|T_0\rangle$.